# Fundamentals of Semantic Web Technologies in medical environments: a case in breast cancer risk estimation

**Iker Huerga**
Research and Development Department
LINKatu, S.L.
Vitoria (Araba), Spain
ihuerga@linkatu.net

**Ainhoa Serna**
Computing and Electronics Department, University of Mondragon
Mondragon (Gipuzkoa), Spain
aserna@eps.mondragon.edu

**Jon Kepa Gerrikagoitia**
Computing and Electronics Department, University of Mondragon
Mondragon (Gipuzkoa), Spain
jkgerrikagoitia@eps.mondragon.edu

**Abstract.** Risk estimation of developing breast cancer poses as the first prevention method for early diagnosis. Furthermore, data integration from different departments involved in the process plays a key role. In order to guarantee patient safety, the whole process should be orchestrated and monitored automatically. Support for the solution will be a linked data cloud, composed by all the departments that take part in the process, combined with rule engines.

## 1 Introduction

The Linked Data (LD) (Bizer, C., et al, 2009) refers to a set of principles to publish data on the web. The Linking Open Data[1] project keeps track of data sets that follow the Linked Data principles. Figure 1[2] gives an overview of data sets that already follow these principles (as of September 2010). Purple colored nodes at the bottom right of the figure refer to life science data sets. Actually, a growing penetration of these technologies into the life science community is becoming evident through many initiatives and conferences (e.g. Bio2RDF[3], BioHackathon[4], and SWAT4LS[5]).

Furthermore, all the assistance areas involved in breast cancer early diagnosis and treatment use its own information system. Due to this fact, data integration between departments is a painful task that is achieved manually most of the times. Thus, the application of a common vocabulary, ontology, to interchange data between departments as well as representing information following Linked Data principles play a key role in this scenario.

In this paper we discuss the benefits of exploiting these technologies to support breast cancer risk estimation and early diagnosis in a real scenario, the Cruces Hospital, in Bilbao, Spain.

---

[1] http://esw.w3.org/SweoIG/TaskForces/CommunityProjects/LinkingOpenData
[2] Figure taken from http://richard.cyganiak.de/2007/10/lod/lod-datasets_2010-09-22_colored.html
[3] http://bio2rdf.org/
[4] http://hackathon3.dbcls.jp/
[5] http://www.swat4ls.org/

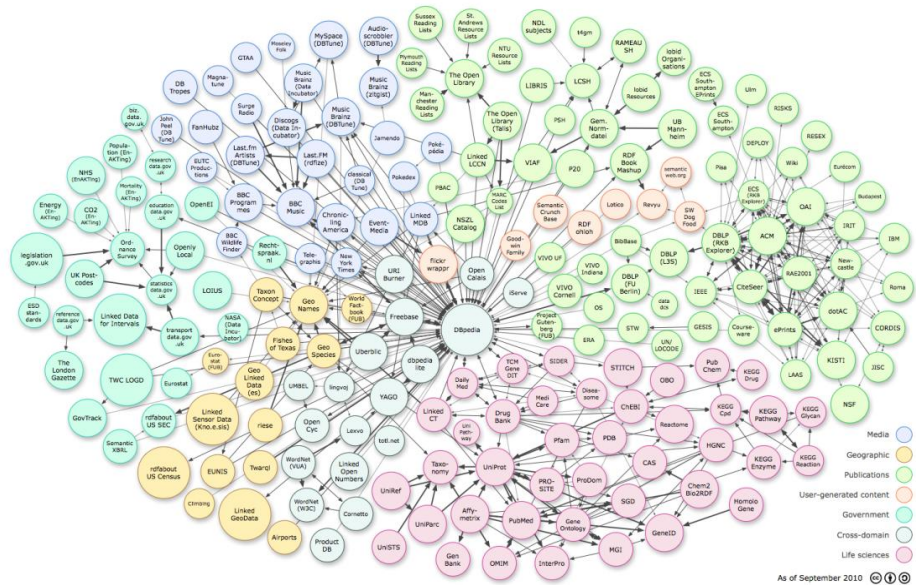

*Figure 1. Linked Open Data cloud as of September 2010*

## 2  Estimating breast cancer risk

A risk factor is anything that increases your chance of developing breast cancer and a protective factor is anything that reduces your risk of developing breast cancer. Some risk factors can be controlled by the patient. For example, if she is overweight, excess pound losses may reduce patient´s risk of developing breast cancer. But other factors are beyond their control. For instance, patients cannot change their gender. Women are much more likely than men to have breast cancer (Giordano et al., 2003). This is mostly because women have more estrogen and progesterone in their bodies. These hormones stimulate breast cell growth, both normal and abnormal. Also, aging is the biggest risk factor for breast cancer, and you cannot stop growing older.

In order to detect the risk factors identified by the current highest level of clinical evidence, a high quality systematic review of the literature has been carried out. The most significant factors identified were genetic inheritance (Kauff et al. 2005), ethnicity (Esther M. John et al., 2007), menstrual history (Clemons et al. 2006), late pregnancy or no pregnancy and breastfeeding (Helewa et al. 2002), even risk factors such as alcohol (Hamajima et al., 2002) or smoking (Collishaw et al, 2009).

Regarding the design of the risk estimating system, a Bayes' theorem iterations approach has been followed.

$$P(Ai/B) = \frac{P(B/Ai)P(Ai)}{P(B)}$$

## 3  Knowledge Representation used in the process

A common semantic vocabulary, ontology, is needed in order to achieve the goals of this approach, estimating the risk of developing breast cancer and interchanging data between the different departments involved in the process. On the one hand, is necessary to semantically annotate the different risk factors due to the population profiles and risks evolve over the time with different prominent risk factors. On the other hand, a common vocabulary should be shared between different department´s information systems in order to automatically interchange information.

Among the different ontologies developed for breast cancer we will focus on the following: cancer ontology at TONES[6] repository and the ontology developed by Mathew Hardy Williams[7] (Hardy, 2008). This study combines both ontologies to produce the ontology used in this research to support the breast cancer department from Cruces Hospital.

Virtuoso Universal Server[8], in its open source edition, storage engine has been used to perform the knowledge base for this study. This storage engine lets us add different semantically annotated data within each RDF Graph. Usage of RDF Graphs is necessary for this kind of application because huge amounts of data are managed by the server.

## 4  Running the application

Taking a typical health record stored on any Hospital's database as starting point, the system will gather the information available in each department involved in the process to enrich the record. After this, a set of inference rules to ascertain the risk of developing breast cancer, which is the main aim of this research, will be applied to obtain the suitable monitoring for the patient.

After these and others rules have been applied, patient's real risk is diagnosed. In case the risk had reached the threshold, patient´s doctor had immediately received an alarm indicating that one of his patients is in high risk of developing breast cancer.

## 5  Future work

Since literature is subject to change, and a systematic review is a hard task, a tool to automatically distinguish the current medical evidence from the noise in the data will be developed. In order to obtain these results SPARQL queries must be done over different endpoints shown in Figure 1. A Linking Transversal (Hartig, O., et al. 2009) approach will be followed to achieve this goal. Using SQUIN[9] service can be an option in order to retrieve the most suitable medical tests to monitor the patient.

---

[6] http://owl.cs.manchester.ac.uk/repository/download?ontology=http://pellet.owldl.com/ontologies/cancer_cc.owl

[7] http://eprints.ucl.ac.uk/15780/1/15780.pdf

[8] http://www.openlinksw.com/wiki/main/Main

[9] http://squin.sourceforge.net/

# References

1. Bizer, C., Heath, T., Berners-Lee,T. (2009). Linked Data - The Story So Far. International Journal on Semantic Web and Information Systems (IJSWIS), Vol. 5(3), Pages 1-22. DOI: 10.4018/jswis.2009081901

2. Clemons, M., Simmons, C.. Identifying menopause in breast cancer patients: considerations and implications. Division of Medical Oncology/Haematology, Princess Margaret Hospital, Suite 5-205, 610 University Avenue, M5G 2M9Toronto, ON, Canada

3. Collishaw NE, et al. Canadian expert panel on tobacco smoke and breast cancer risk. 2009.

4. Esther M. John, PhD; Alexander Miron, PhD; Gail Gong, PhD; Amanda I. Phipps, MPH; Anna Felberg, MS; Frederick P. Li, MD; Dee W. West, PhD; Alice S. Whittemore, PhD. Prevalence of Pathogenic BRCA1 Mutation Carriers in 5 US Racial/Ethnic Groups. JAMA.2007; 298(24):2869-2876.

5. Giordano,S., MD; Aman U. Buzdar, MD; and Gabriel N. Hortobagyi, MD. (2003). Breast Cancer in Men.

6. Hamajima, N., Hirose, K. et al Alcohol, tobacco and breast cancer--collaborative reanalysis of individual data from 53 epidemiological studies, including 58,515 women with breast cancer and 95,067 women without the disease. Br J Cancer 2002; 87; 11; 1234-45.

7. Hardy Williams, M. (2008). Integrating Ontologies and argumentation for decision-making in breast cancer. Doctoral Thesis, University College London.

8. Hartig, O., Bizer, C., Freytag, J.C. (2009) Executing SPARQL Queries over the Web of Linked Data. International Semantic Web Conference, 2009

9. Helewa, M., FRCSC,Winnipeg, Pierre Lvesque, Diane Provencher. (2002). On breast cancer, pregnancy, and breastfeeding

10. Kauff D., Nandita Mitra, Mark E. Robson, Karen E. Hurley, Shaokun Chuai, Deborah Goldfrank, Eve Wadsworth, Johanna Lee, Tessa Cigler, Patrick I. Borgen, Larry Norton, Richard R. Barakat, Kenneth Of. Risk of Ovarian Cancer in BRCA1 and BRCA2 Mutation-Negative Hereditary Breast Cancer Families. Journal of the National Cancer Institute, Vol. 97, No. 18, September 21, 2005